\documentstyle[12pt]{article}

\input{psfig}
\small
\begin{document}
\begin{titlepage}

\title{
{\bf On the Skyrme model prediction for the $N$-$N$ spin-orbit force }}
\author{ {\bf Abdellatif  Abada}\thanks{e-mail: abada@a13.ph.man.ac.uk} \\
{\normalsize 
{\it Theoretical Physics Group, Department of Physics and Astronomy}}\\
{\normalsize {\it University of Manchester, Manchester M13 9PL, UK}} }

\date{}
\maketitle
\begin{abstract}
In the framework of the product ansatz as an approximation for the 
two-baryon system we review in details the derivation of the isoscalar 
nucleon-nucleon spin orbit potential coming from the sixth order term of the 
extended Skyrme model. 
We show that the sixth order term contributes with a positive sign, 
as is the case for the Skyrme term, contrary to the claims of Riska and
Schwesinger. Those authors considered only one part of the force due
to the sixth 
order term and omitted the second part which turns out to be the dominant one. 
Our result is independent of the parameters of the model.
\end{abstract}



\vskip0.5cm
\noindent
MC/TH 96/01 \hfill{January 1996}
\end{titlepage}

There have been several attempts \cite{NR,ASW} to extract the $N$-$N$ 
spin-orbit interaction
from the standard Skyrme Lagrangian \cite{Sk} which includes the non-linear 
$\sigma$ model supplemented with a stabilizing fourth-order term in powers
of the derivatives of the pion field. The calculations were based on 
the product ansatz for the two-baryon system as suggested 
by Skyrme \cite{Sk2}. The advantage of this approximation is that,
beyond its relative simplicity  
as compared to other two-baryon field configurations which can be 
found in the literature, it becomes exact for large $N$-$N$ 
separation\footnote{The region of validity of the product ansatz corresponds 
to a relative distance $r$ larger than 1 fm}. Therefore one can expect the
asymptotic behaviour of the $N$-$N$ spin-orbit force derived from the 
Skyrme model by using the product ansatz agrees with the phenomenology. 
Unfortunately this is not the case for the isoscalar component of the 
spin-orbit force.
All the authors who worked on that subject agree on the result that the
standard Skyrme model predicts an isospin independent spin-orbit force 
with the {\it wrong} sign. Namely, it predicts a {\it repulsive} interaction 
while the phenomenological Bonn potential \cite{Bonn} as well as the Paris
potential \cite{Paris} give an {\it attractive} one. A natural 
reaction that one might have in order to cure this illness, is to improve 
the Skyrme model. Indeed, it has recently been shown \cite{Mo,AM} through 
the study of some properties of the nucleon, that in order to describe
properly low-energy hadron physics one should not restrict oneself
to the standard Skyrme model but consider extensions of this model including 
higher order terms in powers of the derivatives of the pion field. 
Expressed in terms of an SU(2) matrix $U$ which characterizes the pion field,
a 6th-order term corresponding to $\omega$-meson exchange~\cite{Jac},
\begin{equation} \label{L6}
\displaystyle {\cal L}_6 =  - \frac{\beta_{\omega}^2}{2m_{\omega}^2}
B_{\mu}(U) B^{\mu}(U)~,
 \end{equation}
where $B^{\mu} =  \epsilon ^{\mu \nu \alpha \beta} \hbox {Tr} \left(~
(\partial_{\nu} U) U^+ (\partial_{\alpha} U) U^+ (\partial_{\beta} U) U^+
~\right) /24 \pi^2$
is the baryon current \cite{Sk}, $m_{\omega}$ the 
$\omega$-meson mass and  $\beta_{\omega}$ a dimensionless 
parameter related to the $\omega \to \pi \gamma $ width, 
might be a good candidate to solve the problem of the $N$-$N$ isoscalar 
spin-orbit force. 
Riska and Schwesinger~\cite{RS} at first and  K\"albermann and
Eisenberg~\cite{KE} more recently examined the influence of such term on the 
spin orbit interaction\footnote{A dilaton field has also been included 
in \cite{KE}.}. These authors claimed that the inclusion of the
sixth-order term leads to the correct sign (attractive interaction) for the
isoscalar spin-orbit potential. However they considered only one part 
of the interaction due to the sixth-order term in their calculations and 
omitted the second part which arises from the exchange current \cite{RS,NR2}. 
The aim of this paper is to display the derivation of the isoscalar spin-orbit
potential and show that the term omitted in \cite{RS} contributes 
significantly~to~that~force. 

For a system of two interacting solitons, Skyrme \cite{Sk2} suggested 
the use of the product ansatz. Rotational dynamics are also introduced 
to obtain the appropriate spin and isospin structure~\cite{Ad1}. 
Thus, the field configuration of the two-nucleon system separated by 
a vector ${\bf r}$ can be written 
\begin{equation} \begin{array}{cc}\label{pro}
U(A_1,A_2,{\bf x}, {\bf r}) = U_1 U_2 =
 A_1 U_H( {\bf r}_1)A_1^+A_2 U_H( {\bf r}_2) A_2^+ ~~,\\
{\bf r}_1 = {\bf x}-{\bf r}/2 ~~,~~~ 
{\bf r}_2 = {\bf x}+{\bf r}/2~~,
\end{array} \end{equation}
where  $A_1$ and $A_2$ are SU(2) matrices.
To carry out a simultaneous quantization of the relative motion of the two 
nucleons and the rotational motion
we need to treat ${\bf r}, A_1$ and $A_2$ as collective coordinates. Hence
we make all these parameters (${\bf r}, A_1, A_2$) time dependent.
In Eq. (\ref{pro}) 
$U_H$ is the commonly used SU(2) matrix for a single soliton with the 
hedgehog ansatz:
\begin{equation} \label{hed}
U_H({\bf x}) = {\displaystyle \exp[~i\mbox{\boldmath $\tau$}.\hat {\bf x} 
F(\vert {\bf x} \vert) ~] ~~,~~
\hat {\bf x} \equiv {\bf x}/ \vert {\bf x} \vert}~~ ,
\end{equation}
where $F(\vert {\bf x} \vert)$ obeys the usual boundary conditions for 
winding number one, and the $\tau_{a}$'s are the Pauli matrices. 
Let us observe that the chiral angle $F$, which will be used below in the 
numerical calculations, is obtained by solving the static Euler-Lagrange 
equation derived from the extended Skyrme Lagrangian including the non-linear
$\sigma$ model, the 4th-order term, the 6th-order term as well as 
a small chiral symmetry breaking term \cite{AM}.

The spin-orbit potential will emerge due to a coupling between the relative
motion and the spins of the two nucleons so that we have 
to calculate the kinetic energy corresponding to (\ref{L6}). As it is given in
Ref. \cite{RS} it reads
\begin{equation} \label{K6}
K_6({\bf r})= -\frac{\beta_{\omega}^2}{2m_{\omega}^2}
\int \hbox {d}^3 x ~{\bf B}^2(U_1U_2) ~~,
\end{equation}
where ${\bf B}$ is the spatial-component of the baryon current defined after
Eq. (\ref{L6}). 
Before going further, a word of caution should be given here.
In principle before identifying and extracting the spin-orbit potential 
one has to treat carefully the conversion from velocities (orbital and 
rotational) to canonical momenta ( orbital momentum and spin). Namely
one has to start from a Lagrangian formalism, consider the 
 ``classical'' kinetic energy (which is the opposite of Eq.~(\ref{K6})
in the case of  (\ref{L6})), calculate the mass matrix and then
invert it properly in order to move to a Hamiltonian formalism \cite{ASW}.
However for a large relative distance $r$, the region of validity of the 
product ansatz, this procedure \cite{ASW} is equivalent to that of Refs. 
\cite{NR,RS} which we will use here~\cite{Ris}. In the latter one starts
from Eq.~(\ref{K6}), in our case, and make the usual 
identifications~\cite{Ad1,Nym}
\begin{equation} \label{subst}
\displaystyle
\dot {\bf r}_n \to \frac{{\bf p}^{(n)}}{M} ~~,~~ 
 \mbox{\boldmath $\omega$}_n = -\frac{i}{2}
\hbox{Tr} ( \mbox{\boldmath $\tau$}
 A_n^+\dot A_n)  \to \frac{{\bf s}^{(n)}}{2\lambda}~~,~~ n=1,2 ~~,
\end{equation}
where ${\bf p}^{(n)}$ and  ${\bf s}^{(n)}$ are respectively the radial
momentum and the spin of the $n$-th nucleon while
$M$ and $\lambda$ are the mass and the moment of inertia 
of the single soliton.

Inserting the product ansatz (\ref{pro}) into the spatial components of the 
baryon current gives
\begin{equation}\label{bu1u2}
{\bf B}(U_1U_2)= {\bf B}(U_1)+{\bf B}(U_2)+{\bf B}_{\rm ex}(U_1,U_2)~~,
\end{equation}
where ${\bf B}(U_n)$ and ${\bf B}_{\rm ex}(U_1,U_2)$ are the single baryon 
current and the exchange current densities, respectively. The single baryon 
current is given by
\begin{equation} \label{sing}
\displaystyle {\bf B}(U_n)= B_0(\vert{\bf r}_n\vert)
\left(\frac{{\bf p}^{(n)}}{M} +
{\bf r}_n \times \frac{{\bf s}^{(n)}}{\lambda}\right) ~~\hbox {for}~~n=1,2~~,
\end{equation}
where  $B_0(r)= -F'(r)\sin^2F(r)/2\pi^2r^2$ is the baryon density, while the  
rather complicate expression of ${\bf B}_{\rm ex}$ reads
\begin{equation} \label{Bex}
B_{i~{\rm ex}} = \displaystyle
D_{pq}(C) ~\left({\cal A}_{ipql}~\frac{s^{(1)}_l}{2\lambda} + 
{\cal A}_{ipql}'~  \frac{p^{(1)}_l}{M} + {\cal B}_{ipql}~
\frac{s^{(2)}_l}{2\lambda} +   
{\cal B}_{ipql}'~  \frac{p^{(2)}_l}{M} \right) ~~,
\end{equation}
in which  $C=A_1^+A_2$ and $D_{pq}$ is the $3\times3$ rotation matrix in the 
adjoint representation 
\begin{equation} \label{iso}
D_{pq}(C)= \frac {1}{2} \hbox {Tr} (\tau_p C \tau_q C^+)~~. 
\end{equation}
The sum from 1 to 3 on repeated indices in Eq. (\ref{Bex}), and from here on, 
is understood.
The four tensors appearing in the expression of ${\bf B}_{\rm ex}$ depend only
on the positions of the two nucleons. They read 
\begin{equation}\begin{array}{ll} \label{ABAB}
{\cal A}_{ipql}({\bf r}_1,{\bf r}_2) = &{\displaystyle
\left(2 \epsilon_{ijk}\epsilon_{abp}~T^{(1)}_{lb} ~R^{(1)}_{ja} ~R^{(2)}_{qk}
- T^{(1)}_{lp}~\alpha^{(2)}_{iq} \right)/4\pi^2 ~~,}\\
{\cal A}_{ipql}'({\bf r}_1,{\bf r}_2) = &{\displaystyle
\left(2 \epsilon_{ijk}\epsilon_{abp}~R^{(1)}_{lb} ~R^{(1)}_{ja} ~R^{(2)}_{qk}
- R^{(1)}_{lp}~\alpha^{(2)}_{iq} \right)/4\pi^2 ~~,}\\
{\cal B}_{ipql}({\bf r}_1,{\bf r}_2) = &{\displaystyle
\left(-2 \epsilon_{ijk}\epsilon_{abq}~T^{(2)}_{bl} ~R^{(1)}_{kp} ~R^{(2)}_{aj}
+ T^{(2)}_{ql}~\alpha^{(1)}_{pi} \right)/4\pi^2 =
-{\cal A}_{iqpl}(-{\bf r}_2,-{\bf r}_1)~~,}\\
{\cal B}_{ipql}'({\bf r}_1,{\bf r}_2) = &{\displaystyle
\left(2 \epsilon_{ijk}\epsilon_{abq}~R^{(2)}_{bl} ~R^{(1)}_{kp} ~R^{(2)}_{aj}
- R^{(2)}_{ql}~\alpha^{(1)}_{pi} \right)/4\pi^2 =
{\cal A}_{iqpl}'(-{\bf r}_2,-{\bf r}_1)~~,}
\end{array}\end{equation}
where
\begin{equation}\begin{array}{ll} \label{defs}
&{\displaystyle R^{(n)}_{ia} =-\frac{i}{2} \hbox {Tr} (\tau_a U_H^+({\bf r}_n) 
\partial_i U_H({\bf r}_n) ) =
 r_n \left( c_ns_n(\delta_{ia} -  \hat {\bf r}_{n i} 
\hat {\bf r}_{n a} ) + \frac{F_n'}{r_n} \hat {\bf r}_{n i} \hat {\bf r}_{n a}
+ s_n^2 \epsilon_{ial} \hat {\bf r}_{n l} \right)}~, \\
&{\displaystyle T^{(n)}_{bc} = 2~\epsilon_{ilb}~r_{n l}~R^{(n)}_{ic}
= 2~r_n^2\left(-s_n^2 (\delta_{bc} -  \hat {\bf r}_{n b} \hat {\bf r}_{n c} ) 
+ s_n c_n ~\epsilon_{b c l}~\hat {\bf r}_{n l} \right) }~,\\
&{\displaystyle 
\alpha^{(n)}_{ia}=\epsilon_{ijk}~\epsilon_{abc}~R^{(n)}_{bj}~R^{(n)}_{ck} }~,\\
&{\displaystyle 
r_n \equiv \vert {\bf r}_n \vert ~,~~  F_n \equiv F(r_n) ~,~~ 
F'_n \equiv \frac {\partial F_n}{\partial r_n} ~,~~
c_n \equiv \frac {\cos F_n}{r_n} ~,~~ s_n \equiv 
\frac{\sin F_n}{r_n} ~,~~~~n = 1,2~~.}
\end{array}\end{equation}
By inserting now the expression (\ref{bu1u2}) in the kinetic energy (\ref{K6})
we obtain 
\begin{equation}\label{K6p}
K_6({\bf r})= -\frac{\beta_{\omega}^2}{2m_{\omega}^2}
\int \hbox {d}^3 x ~\left(2{\bf B}(U_1){\bf .B}(U_2) + 
{\bf B}_{\rm ex}^2 + 2~({\bf B}(U_1)+{\bf B}(U_2)~)~{\bf .B}_{\rm ex} ~
+ ~\cdots \right)~~,
\end{equation}
where we omitted the terms corresponding to the sum of the squares of the 
single baryon currents since they do not bring any coupling between the two 
nucleons and thus do not contribute to the spin-orbit force.
From Eqs. (\ref{Bex}) and (\ref{iso}) one sees that the exchange current
${\bf B}_{\rm ex}$ contains the isospin factor 
$\mbox{\boldmath $\tau$}^{(1)}.\mbox{\boldmath $\tau$}^{(2)}$ due to the 
projection theorem \cite{WW} 
\begin{equation}\label{proj}
\langle N_1'N_2'\vert D_{ab}(C)\vert N_1N_2\rangle = \frac{1}{9}
(\mbox{\boldmath $\tau$}^{(1)}.\mbox{\boldmath $\tau$}^{(2)})
\sigma_a^{(1)}\sigma_b^{(2)}~~,
\end{equation}
while the single baryon currents are isospin independent [cf., 
Eq. (\ref{sing})~]. Therefore the isoscalar component of the spin-orbit force 
arises only from the first and the second term in the expression~(\ref{K6p}).
The calculation of the spin-orbit potential derived from the first term is 
straightforward. By inserting the definition of the single currents 
(\ref{sing}) and keeping the terms proportional to ${\bf L.S}$ where 
${\bf S}={\bf s}^{(1)}+{\bf s}^{(2)}$ is the total spin and 
${\bf L}={\bf r}\times {\bf p}$, is the angular momentum,  
${\bf p}$ being the relative momentum, i.e.,  
${\bf p}={\bf p}^{(2)}=-{\bf p}^{(1)}$, we obtain
\begin{equation}\label{riska-term0}
- \frac{\beta_{\omega}^2}{2m_{\omega}^2}
\int \hbox {d}^3 x ~2{\bf B}(U_1){\bf .B}(U_2) ~\to 
  \frac{\beta_{\omega}^2}{2m_{\omega}^2} \frac{1}{M\lambda}
\Sigma_6(r) ~{\bf L.S}~~,
\end{equation}
where 
\begin{equation}\label{riska-term}
\displaystyle \Sigma_6(r) =
-\int \hbox {d}^3 x ~ B_0(r_1)B_0(r_2) ~=  
 -\frac{1}{4\pi^4}\int {\rm d}^3x~ s_1^2s_2^2 F_1' F_2' ~~.
\end{equation}
The above formulae are equivalent to that obtained in Ref. \cite{RS} and the 
function $\Sigma_6$ as defined in Eq. (\ref{riska-term})
is obviously negative (see Fig.1).

The second term in the kinetic energy (\ref{K6p}) has been omitted in the 
calculations of Ref.~\cite{RS}. In fact in their paper Riska and Schwesinger 
\cite{RS} referred to the article \cite{NR2} for the expression of the 
exchange baryon current. However in that paper \cite{NR2} the momentum 
dependent terms, as given here in Eq. (\ref{Bex}), have been dropped from 
the exchange current, 
since their contribution is of less significance than the spin dependent 
terms for the calculation of the deuteron form factors, which was 
the purpose of the article \cite{NR2};
and thus no spin-orbit coupling will arise from their incomplete exchange 
current \cite{RS}.
This is obviously not the case when one takes into consideration the entire 
expression (\ref{Bex}). Even though the exchange current (\ref{Bex}) is
proportional to the rotation matrix $D_{pq}$, and thus to 
$\mbox{\boldmath $\tau$}^{(1)}.\mbox{\boldmath $\tau$}^{(2)}$, 
${\bf B}_{\rm ex}^2$  contains an isoscalar component. Indeed one has the 
formula
\begin{equation} \label{DD}
\displaystyle D_{pq}(C) D_{p'q'}(C) = \frac{1}{3} \delta_{pp'}\delta_{qq'} 
+\frac{1}{2} \epsilon_{pp'p''}~\epsilon_{qq'q''}~D_{p''q''}(C) + 
\sum_{mn} C_{pqp'q'}^{mn} {\cal D}_{mn}^{j=2}(C) ~~,
\end{equation}
which can be obtained by expressing the $D_{pq} (C)$ in terms of the Wigner
${\cal D}$-functions.
By replacing now $D_{pq}(C) D_{p'q'}(C)$ by 
$\frac{1}{3} \delta_{pp'}\delta_{qq'}$ (we are indeed interested only in the 
isoscalar part of the spin-orbit force) in the expression of 
${\bf B}_{\rm ex}^2$ and keeping only the relevant terms which might 
generate an ${\bf L.S}$ contribution we obtain
\begin{equation}\label{exch-term}
- \frac{\beta_{\omega}^2}{2m_{\omega}^2}
\int \hbox {d}^3 x ~{\bf B}_{\rm ex}^2 ~\to ~
 \frac{\beta_{\omega}^2}{2m_{\omega}^2} \frac{1}{M\lambda}~S_l 
 \left(-\frac{1}{3} \int \hbox {d}^3 x ~ 
{\cal A}_{ipql} ( {\cal B}'_{ipql'} - {\cal A}'_{ipql'})~\right) p_{l'}~~.
\end{equation}

After some tedious calculations\footnote{These have been checked by
computer algebra.} [cf. Eqs. (\ref{ABAB},\ref{defs}) ], the term 
between brackets in the above equation turns out to have the structure~
$\Sigma_{6 {\rm ex}}(r) ~\epsilon_{il'l}~r_i $~
so that one obtains
\begin{equation}\label{exch-term2}
- \frac{\beta_{\omega}^2}{2m_{\omega}^2}
\int \hbox {d}^3 x ~{\bf B}_{\rm ex}^2 ~\to ~
 \frac{\beta_{\omega}^2}{2m_{\omega}^2} \frac{1}{M\lambda}~
\Sigma_{6 {\rm ex}}(r) ~{\bf L.S} ~~.
\end{equation}
The expression of $\Sigma_{6{\rm ex}}$ reads
\begin{equation}\begin{array}{rr}\label{sig6ex}
 \displaystyle 
\Sigma_{6{\rm ex}}=  \frac{1}{3\pi^4r}
\int {\rm d}^3 x~ r_1s_1^2\left(
 \frac{1}{4} \hat {\bf r.}\hat {\bf r}_1~ 
(\hat {\bf r}_1\hat {\bf .r}_2)^2 (s_1^2-F_1^{'2}) (s_2^2-F_2^{'2}) +
\frac{1}{4} \hat {\bf r.}\hat {\bf r}_2~
\hat {\bf r}_1\hat {\bf .r}_2  (F_2^{'2}-s_2^2) \times
\right.\\
\displaystyle 
\left. (2F_1^{'2}+s_1^2+s_2^2) ~+
\hat {\bf r.}\hat {\bf r}_1 ~(~ \frac{3}{2} s_2^2 F_1' F_2' -
\frac{3}{4} F^{'2}_1F^{'2}_2 - \frac{1}{2}s_1^2F^{'2}_2 - \frac{3}{4}
s_2^2F^{'2}_1 - \frac{7}{4} s_2^2F^{'2}_2 - \frac{3}{4}s_2^4 -s_1^2s_2^2 )~
\right)
\end{array} \end{equation}
We refer to Eqs. (\ref{hed},\ref{defs}) for the notations used in the above 
formula.
Thus the isospin independent spin-orbit force generated by the 6th-order 
term  in powers of the derivatives of the pion field~(\ref{L6}) reads 
\begin{equation}\label{SO}
\displaystyle
V_{\rm SO} \equiv \frac{\beta_{\omega}^2}{2m_{\omega}^2} \frac{1}{M\lambda}~
\Sigma_{6\rm tot}(r)~ {\bf L.S} =
\frac{\beta_{\omega}^2}{2m_{\omega}^2} \frac{1}{M\lambda}~\left(
\Sigma_{6}(r) + \Sigma_{6{\rm ex}}(r)\right)~{\bf L.S} ~~,
\end{equation}
where the expressions of $\Sigma_6(r)$ and  $\Sigma_{6{\rm ex}}(r)$ are given 
in Eqs.~(\ref{riska-term}) and (\ref{sig6ex}) respectively. In Fig.1 we plot 
these two functions, with respect 
to the relative distance~$r$. We see from that figure that $\Sigma_6(r)$ is 
negative, as it was expected and found in \cite{RS}. However the contribution 
of the exchange current which has been omitted in \cite{RS}, 
$\Sigma_{6{\rm ex}}$, is positive and 
larger than  $\Sigma_6(r)$ so that the total function $\Sigma_{6\rm tot}$,
also displayed in Fig.1, is always positive. Therefore the 6th-order term
generated by $\omega$-meson exchange (\ref{L6}) gives a {\it repulsive } 
isoscalar spin-orbit interaction, contrary to the claims of Ref. \cite{RS}.  
Our result is independent of the parameters $\beta_{\omega}, m_{\omega}$ 
as one can see from Eq. (\ref{SO}). Of course the chiral angle $F$ on which 
the functions $\Sigma$ depend, depends implicitly on the parameters of the 
extended Skyrme model \cite{Sk,AM} ($f_{\pi}, e, \beta_{\omega}$). 
Nevertheless 
the main result of this paper, namely, the dominance of the positive 
contribution $\Sigma_{6 {\rm ex}}$ over the negative one $\Sigma_6$, is 
independent of the parameters of the model. We checked this result numerically 
by considering several sets of parameters which can be found in the 
literature; realistic as well as unrealistic ones.

\begin{figure}[t]
\centerline{\psfig{figure=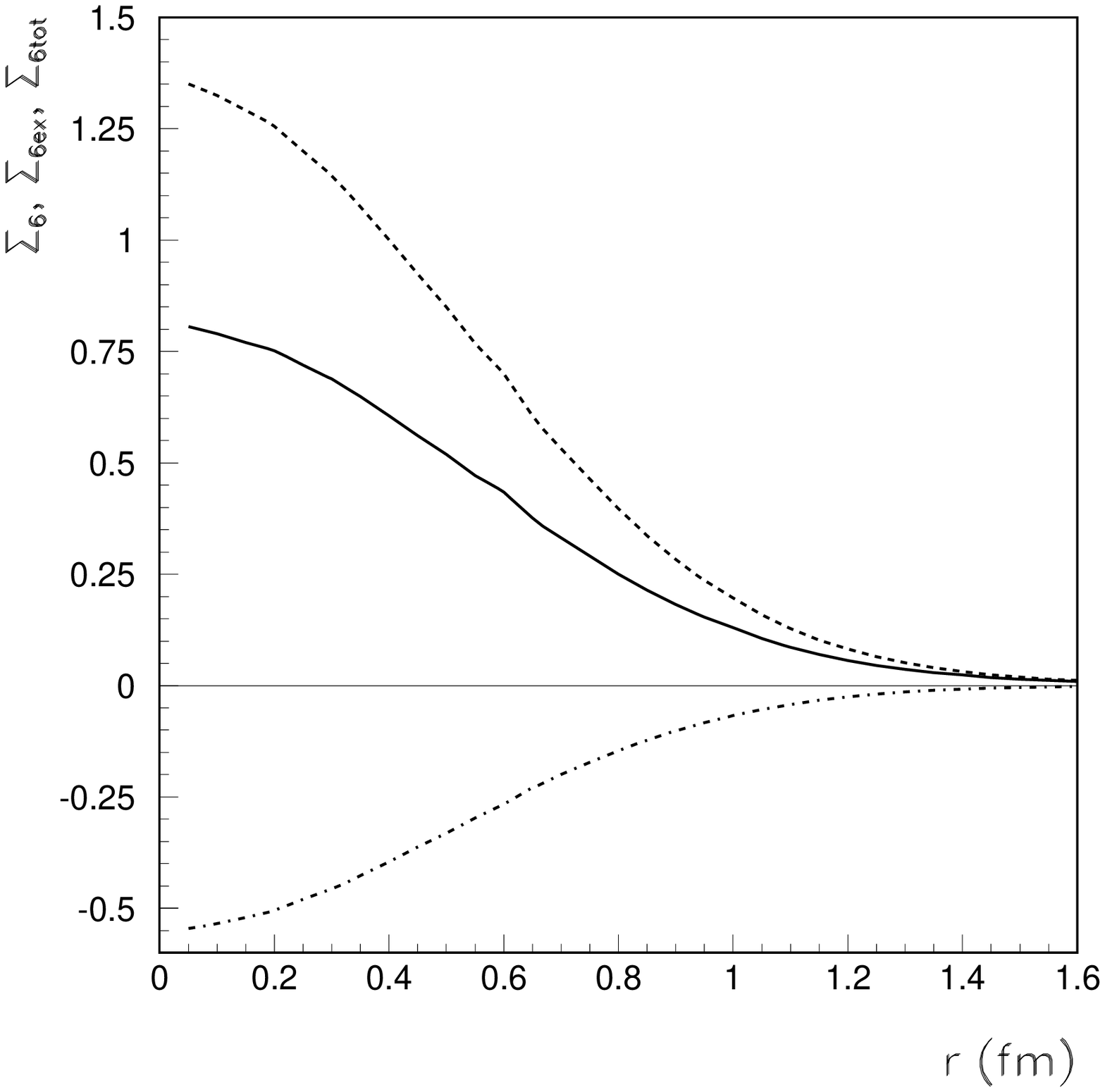,height=10cm,width=14cm}}
\begin{center}{
\parbox{15cm}{ {\large {\bf FIG.~1}} ~:
{\footnotesize
The functions $\Sigma_6$ (dashed-dotted line), $\Sigma_{6{\rm ex}}$ 
(dashed line) and  $\Sigma_{6{\rm tot}}$ (full line) in~fm$^{-3}$, as given in 
Eqs. (\ref{riska-term},\ref{sig6ex},
\ref{SO}), with respect to the relative $N$-$N$ distance $r$.} }
}\end{center}
\end{figure}

In this letter we derived in details the isospin-independent spin-orbit force
from the sixth-order order term in powers of the derivatives of the pion 
field (\ref{L6}). By taking into account the results of Refs. \cite{NR,ASW}
concerning the Skyrme term \cite{Sk}, we arrive to our major
conclusion: neither the Skyrme term which can be
derived from a local approximation of an effective $\rho$ model,
nor the sixth-order term generated by $\omega$-meson exchange,
reproduces the phenomenological isoscalar $N$-$N$ spin-orbit interaction. 
Indeed both terms give a {\it repulsive} force while it should be 
{\it attractive} according to \cite{Bonn,Paris}. 
Hence the problem of the sign for this force is still unsolved
 and one should try to understand why. The key to the
problem is not at hand yet but one can suggest some hints. The 
spin-orbit force is a relativistic problem and since we are in a region of 
large distance (the region of validity of the product ansatz) where
this force is weak, any relativistic effects, even small ones, can affect 
that interaction. Then one should consider such relativistic effects in the 
calculations (we are thinking, e.g., to the Thomas precession effects due to 
the rotation of the nucleons).  
Also one sees from the Bonn potential \cite{Bonn} that the scalar
degrees of freedom, namely the $\sigma$-meson which can be viewed as 
the responsible for the enhancement of the $\pi\pi~S$-wave, provide  one of 
the major contributions to the isoscalar spin-orbit force. 
So maybe one has  to investigate the spin-orbit part of the two-pion 
exchange potential within the Skyrme model \cite{Vin} in order to correct 
the anomaly of that sign. This idea to account for a scalar field has been 
investigated in Ref. \cite{KE} in which  a dilaton field is coupled to 
Skyrmions in order to mimic the scale breaking of QCD. Even though it has been 
pointed out in Ref. \cite{Mk} that a dilaton field is not suitable to provide
a good description of low-energy hadron physics, the fact remains that 
a combination of the sixth-order term (\ref{L6}) and dilaton coupling, 
i.e., {\it scalar} meson coupling, yields an attractive isoscalar spin-orbit 
force as it has been claimed in \cite{KE}. 
However this result remains questionable since these authors, as those of
Ref. \cite{RS}, did not take into account the exchange current contribution.

\vskip .5cm
\noindent{\large {\bf Acknowledgments} }

\noindent We express our deep appreciation to  M. C. Birse for helpful 
discussions. We also thank him as well as J. McGovern  for a critical 
reading of the manuscript.


\begin{thebibliography}{99}
\bibitem{NR} E. M. Nyman and D. O. Riska, Phys. Lett. B {\bf 175} (1986) 392 ;
{\bf 183} (1987) 7 ; 
D. O. Riska and K. Dannbom, Phys. Scr. {\bf 37} (1988) 7 ; 
T. Otofuji {\it et al}, Phys. Lett. B {\bf 205} (1988) 145.
\bibitem{ASW} R. D. Amado, B. Shao and N. R. Walet, Phys. Lett. B {\bf 314} 
(1993) 159 and an erratum : Phys. Lett. B {\bf 324} (1994) 467 ;\\
B. Shao, N. R. Walet and R. D. Amado, 
Phys. Rev. C {\bf 48} (1993) 2498 and an erratum :  
Phys. Rev. C {\bf 49} (1994) 3360.
\bibitem{Sk} T. H. R. Skyrme, Proc. Roy. Soc. {\bf A260} (1961) 127.
\bibitem{Sk2} T. H. R. Skyrme, Nucl. Phys. {\bf 31} (1962) 556.
\bibitem{Bonn}  R. Machleidt, K. Holinde and Ch. Elster, Phys. Rep. {\bf 149} 
(1987) 1.
\bibitem{Paris} M. Lacombe {\it et al}, Phys. Rev. C{\bf 21} (1980) 861.
\bibitem{Mo} B. Moussallam, Ann. Phys. (N.Y.) {\bf 225} (1993) 264.
\bibitem{AM} A. Abada and H. Merabet, Phys. Rev. D {\bf 48} (1993) 2337.
\bibitem{Jac} A. Jackson {\it et al}, Phys. Lett. B {\bf 154} (1985) 101.
\bibitem{RS} D. O. Riska and B. Schwesinger, Phys. Lett. B {\bf 229} (1989) 
339.
\bibitem{KE} G. K\"albermann and J. M. Eisenberg, Phys. Lett. B {\bf 349}
(1995) 416.
\bibitem{NR2} E. M. Nyman and D. O. Riska, Nucl. Phys. A {\bf 468} (1987) 473.
\bibitem{Ad1} G. S. Adkins, C. R. Nappi and E. Witten, Nucl. Phys.
{\bf B228} (1983) 552.
\bibitem{Ris} D. O. Riska, private correspondence.
\bibitem{Nym} E. M. Nyman, Phys. Lett. B {\bf 162} (1985) 244.
\bibitem{WW} T. Walhout and J. Wambach, Int. Jou. Mod. Phys. E Vol. 1, No. 4
(1992) 665.
\bibitem{Vin} R. Vinh Mau, private communication; N. R. Walet, private 
communication.
\bibitem{Mk} M. C. Birse, J. Phys. G {\bf 20} (1994) 1287.
\end{thebibliography}
\end{document}